# Towards algebraic methods for maximum entropy estimation

**Ambedkar Dukkipati**

EURANDOM, P.O. Box 513, 5600 MB Eindhoven, The Netherlands

E-mail: `dukkipati@eurandom.tue.nl`

**Abstract.** We show that various formulations (*e.g.,* dual and Kullback-Csiszar iterations) of estimation of maximum entropy (ME) models can be transformed to solving systems of polynomial equations in several variables for which one can use celebrated Gröbner bases methods. Posing of ME estimation as solving polynomial equations is possible, in the cases where feature functions (sufficient statistic) that provides the information about the underlying random variable in the form of expectations are integer valued.

1. Introduction

Algebra has always played an important role in statistics, a classical example being linear algebra. There are also many other instances of applying algebraic tools in statistics (e.g (Viana & Richards 2001)). But, treating statistical models as algebraic objects, and thereby using tools of computational commutative algebra and algebraic geometry in the analysis of statistical models is very recent and has led to the still evolving field of *algebraic statistics*.

The use of computational algebra and algebraic geometry in statistics was initiated in the work of Diaconis and Sturmfels (Diaconis & Sturmfels 1998) on exact hypothesis tests of conditional independence in contingency tables, and in the work of Pistone et al. (Pistone et al. 2001) in experimental design. The term 'Algebraic Statistics' was first coined in the monograph by Pistone et al. (Pistone et al. 2001) and appeared recently in the title of the book by Pachter and Sturmfels (Pachter & Sturmbfels 2005).

To extract the underlying algebraic structures in discrete statistical models, algebraic statistics treat statistical models as affine varieties. (An affine variety is the set of all solutions to family of polynomial equations.) Parametric statistical models are described in terms of a polynomial (or rational) mapping from a set of parameters to distributions. One can show that many statistical models, for example independence models, Bernouli random variable etc. (see (Pachter & Sturmbfels 2005) for more examples), can be given this algebraic formulation, and these are referred to as algebraic statistical models.

Exponential models, which form the important class of statistical models, are studied in algebraic statistics under the name 'toric' models by using maximum likelihood methods. Toric models are algebraic statistical models and the term 'toric' comes from important algebraic objects known as 'toric ideals' in computational algebra. In this view of very established role of information theory in statistics (Kullback 1959, Csiszár & Shields 2004) this paper attempts to describe maximum entropy models in algebraic statistical framework.

In particular, we show that maximum entropy models (also minimum relative-entropy models) are indeed toric models, when the functions that provide the information about the underlying random variable in the form of expected values are integer valued. We also show that when the information is available in the form of sample means, by modifying maximum entropy prescriptions calculating model parameters amounts to solving set of polynomial equations. This establishes a fact that set of statistical models results from maximum entropy methods are indeed algebraic varieties.

A note on the results presented in this paper: we will not present the details on Gröbner bases theory and related concepts to solve the polynomial equations due to space constraint; we refer reader to text books on computational algebra and Gröbner basis theory (Adams & Loustaunau 1994, Cox et al. 1991).

We organize our paper as follows. In § 2 we give basic notions of algebra and introduce notation along with an introduction to algebraic statistics. § 3 describes maximum entropy (ME) prescriptions in algebraic statistical framework by introducing important algebraic objects called toric ideals. In § 4 we show how one can transform the problem of calculating ME distributions to solving set of polynomial equations.

## 2. Algebraic Statistical Models

### 2.1. Basic notions of Algebra

Through out this paper $k$ represents a field. A monomial in $n$ indeterminates $x_1, \ldots, x_n$ is a power product of the form $x_1^{\alpha_1} \ldots x_n^{\alpha_n}$, where all the exponents are nonnegative integers, i.e. $\alpha_i \in \mathbb{Z}_{\geq 0}$, $i = 1, \ldots n$. One can simplify the notation for monomial as follows: denote $\alpha = (\alpha_1, \ldots, \alpha_n) \in \mathbb{Z}_{\geq 0}^n$ and by using multi-index notation we set

$$x^\alpha = x_1^{\alpha_1} \ldots x_n^{\alpha_n}$$

with the understanding that $x = (x_1, \ldots, x_n)$. Note that $x^\alpha = 1$ when ever $\alpha = (0, \ldots, 0)$. Once the order of the indeterminates are fixed, monomial $x_1^{\alpha_1} \ldots x_n^{\alpha_n} = x^\alpha$ is identified by $(\alpha_1, \ldots, \alpha_n)$. Hence, set of all monomials in indeterminates $x_1, \ldots, x_n$ can be represented by $\mathbb{Z}_{\geq 0}^n$. Theory of monomials is central to the celebrated Gröbner bases theory in computational algebra which provides tools for solving set of polynomial equations and related problems in algebraic geometry (Mishra & Yap 1989). Monomial theory itself plays important role in algebraic statistics in the representation of exponential models where probabilities are expressed in terms of power products (Rapallo 2006).

A polynomial $f$ in $x_1, \ldots, x_n$ with coefficients in $k$ is a finite linear combination of monomials and can be written in the form of

$$f = \sum_{\alpha \in \Lambda_f} a_\alpha x^\alpha ,$$

where $\Lambda_f \subset \mathbb{Z}_{\geq 0}^n$ is a finite set and $a_\alpha \in k$. The collection of all polynomials in the indeterminates $x_1, \ldots, x_n$ is the set $k[x_1, \ldots, x_n]$ and it has structure not only of a vector space but also of a ring. Indeed the ring structure of $k[x_1, \ldots, x_n]$ plays main role in computational algebra and algebraic geometry.

A subset $\mathfrak{a} \subset k[x_1, \ldots, x_n]$ is said to be ideal if it satisfies: (i) $0 \in \mathfrak{a}$ (ii) $f, g \in \mathfrak{a}$, then $f + g \in \mathfrak{a}$ (iii) $f \in \mathfrak{a}$ and $h \in k[x_1, \ldots, x_n]$ and then $hf \in \mathfrak{a}$. A set $V \subset k^n$ is said to be affine variety if there exists $f_1, \ldots, f_s \in k[x_1, \ldots, x_n]$ such that

$$V = \{(c_1, \ldots c_n) \in k^n : f_i(c_1, \ldots c_n) = 0, 1 \leq i \leq s\} .$$

We use the notation $\mathcal{V}(f_1, \ldots, f_s) = V$.

### 2.2. Algebraic Statistical Model

At the very core of the field of algebraic statistics lies the notion of an 'algebraic statistical model'. While this notion has the potential of serving as a unifying theme for algebraic statistics, there is no unified definition of an algebraic statistical model (Drton

& Sullivant 2006). Here, we adopt the appropriate definition of statistical model from (Pachter & Sturmbfels 2005, Drton & Sullivant 2006). For a recent elaborate discussion on formal definition of algebraic statistical models one can refer to (Drton & Sullivant 2006).

Let $X$ be a discrete random variable taking finitely many values from the set $[m] = \{1, 2, \ldots m\}$. A probability distribution $p$ of $X$ is naturally represented as a vector $p = (p_1, \ldots, p_m) \in \mathbb{R}^m$ if we fix the order on $[m]$. Then set of all probability mass functions (pmfs) of $X$ is called probability simplex

$$\Delta_{m-1} = \{p = (p_1, \ldots, p_m) \in \mathbb{R}^m_{\geq 0} : \sum_{i=1}^{m} p_i = 1\} . \tag{1}$$

The index $m - 1$ indicates the dimension of the simplex $\Delta_{m-1}$. A statistical model $\mathcal{M}$ is a subset of $\Delta_{m-1}$ and is said to be algebraic if $\exists f_1, \ldots, f_s \in k[p_1, \ldots, p_m]$ such that

$$\mathcal{M} = \mathcal{V}(f_1, \ldots, f_s) \cap \Delta_{m-1} .$$

Now we move on to parametric statistical models and their algebraic formulations.

Let $\Theta \subseteq \mathbb{R}^d$ be a parametric space and $\kappa : \Theta \to \Delta_{m-1}$ be a map. The image $\kappa(\Theta)$ is called parametric statistical model. Given a statistical model $\mathcal{M} \subseteq \Delta_{m-1}$, by parametrization of $\mathcal{M}$ we mean, identifying a set $\Theta \subseteq \mathbb{R}^d$ and a function $\kappa : \Theta \to \Delta_{m-1}$ such that $\mathcal{M} = \kappa(\Theta)$. To describe more general statistical models in algebraic framework we need following notion of *semi-algebraic set*.

**Definition 2.1.** *A set $\Theta \subseteq \mathbb{R}^d$ is called semi-algebraic set, if there are two finite collection of polynomials $F \subset k[x_1, \ldots, x_d]$ and $G \subset k[x_1, \ldots, x_d]$ such that*

$$\Theta = \{\theta \in \mathbb{R}^d : f(\theta) = 0, \forall f \in F \text{ and } g(\theta) \geq 0, g \in G\} .$$

Now we have following definition of parametric algebraic statistical model.

**Definition 2.2.** *Let $\Delta_{m-1}$ be a probability simplex and $\Theta \subset \mathbb{R}^d$ be a semi-algebraic set. Let $\kappa : \mathbb{R}^d \to \mathbb{R}^m$ be a rational function (a rational function is a quotient of two polynomials) such that $\kappa(\Theta) \subseteq \Delta_{m-1}$. Then the image $\mathcal{M} = \kappa(\Theta)$ is a parametric algebraic statistical model.*

Conversely, a parametric statistical model $\mathcal{M} = \kappa(\Theta) \subseteq \Delta_{n-1}$ is said to be algebraic if $\Theta$ is semi-algebraic set and $\kappa$ is a rational function. From now on we refer to 'parametric algebraic statistical models' as 'algebraic statistical models'.

In this paper we consider following special case of algebraic statistical models (cf. (Pachter & Sturmbfels 2005, pp 7)). Consider a map

$$\begin{aligned} \kappa : \Theta(\subseteq \mathbb{R}^d) &\to \mathbb{R}^m \\ \kappa : \theta = (\theta_1, \ldots, \theta_d) &\mapsto (\kappa_1(\theta), \ldots, \kappa_m(\theta)) \end{aligned} \tag{2}$$

where $\kappa_i \in k[\theta_1, \ldots, \theta_d]$. We assume that $\Theta$ satisfies $\kappa_i(\theta) \geq 0$, $i = 1, \ldots, m$ and $\sum_{i=1}^{m} \kappa_i(\theta) = 1$ for any $\theta \in \Theta$. Under these conditions $\kappa(\Theta)$ is indeed an algebraic statistical model (Definition 2.2) since $\kappa(\Theta) \subset \Delta_{m-1}$, $\kappa$ is a polynomial function and $\Theta$ is a semi-algebraic set ($H = \{\sum_{i=1}^{m} f_i - 1\}$ and $G = \{f_i : i = 1, \ldots, m\}$ in the Definition 2.1).

Some statistical models are naturally given by a polynomial map $\kappa$ (2) for which the condition $\sum_{i=1}^{m} \kappa_i(\theta) = 1$ does not hold. If this is the case one can consider following algebraic statistical model:

$$\kappa : \theta = (\theta_1, \ldots, \theta_d) \mapsto \frac{1}{\sum_{i=1}^{m} \kappa_i(\theta)} (\kappa_1(\theta), \ldots, \kappa_m(\theta)) , \tag{3}$$

assuming that remaining conditions that have been specified for the model (2) are valid here too. The only difference is that instead of $\kappa$ being a polynomial map, we have it as a rational map.

## 3. ME in algebraic statistical setup

### 3.1. Toric Models

In the algebraic description of exponential models monomials and binomials play a fundamental role. The study of relations of power products lead to the theory of toric ideals in the commutative algebra (Sturmbfels 1996). Here we describe basic notion of toric ideal that are relevant to representation and computation of discrete exponential models; for more details on theory and computation of toric ideals one can refer to (Sturmbfels 1996, Bigatti & Robbiano 2001, Bigatti et al. 1999).

Before we give the definition of toric ideal, we describe the notion of Laurent polynomial. If we allow negative exponents in a polynomial i.e., polynomial of the form $f = \sum_{\alpha \in \Lambda_f} a_\alpha x^\alpha$ where $\alpha \in \mathbb{Z}^n$, it is known as Laurent polynomial ($\Lambda_f \subset \mathbb{Z}_{\geq 0}^n$ is finite). Set of all Laurent polynomials in the indeterminates $x_1, \ldots, x_n$ is denoted by $k[x_1^\pm, \ldots, x_n^\pm]$ and it also has a structure of a ring.

Now we define the toric ideal.

**Definition 3.1.** *Let $A = [a_{ij}] \in \mathbb{Z}^{d \times n}$ be a matrix with rank $d$. Consider the ring homeomorphism*

$$\hat{\pi} : k[x_1, \ldots, x_n] \to k[\theta_1^\pm, \ldots, \theta_d^\pm]$$
$$\hat{\pi} : x_j \mapsto \theta_1^{a_{1j}} \ldots \theta_1^{a_{dj}} \tag{4}$$

*The toric ideal $\mathfrak{a}_A$ of $A$ is defined as the kernel of the map $\hat{\pi}$, i.e., $\mathfrak{a}_A = ker\hat{\pi}$.*

The mapping $\hat{\pi}$ can be viewed as "parametrization" and which can be explained by the following description of $\hat{\pi}$. Consider a map

$$\pi : \mathbb{Z}_{\geq 0}^n \to \mathbb{Z}^d$$
$$\pi : u = (u_1, \ldots, u_n) \mapsto Au. \tag{5}$$

The map $\pi$ lifts to the ring homomorphism $\hat{\pi}$ in the sense of action of $\hat{\pi}$ on $x^u = x_1^{u_1} \ldots x_n^{u_n} \in k[x_1, \ldots, x_n]$. That is

$$\hat{\pi}(x^u) = \hat{\pi}(x_1^{u_1}, \ldots, x_n^{u_n}) = \left(\prod_{i=1}^{d} \theta_i^{a_{i1}}\right)^{u_1} \ldots \left(\prod_{i=1}^{d} \theta_i^{a_{in}}\right)^{u_n} \tag{6}$$

$$= \prod_{i=1}^{d} \theta_i^{\sum_{j=1}^{n} a_{ij} u_j} = \theta^{Au} \ . \tag{7}$$

Toric ideal theory plays an important role in applications of computational algebraic geometry like integer programming etc.(cf. (Sturmbfels 1996)). Note that in the algebraic descriptions of exponential models and their maximum likelihood estimates only non-negative cases of toric ideals (and hence toric models) is considered i.e., the matrix $A = [a_{ij}]$ in Definition 3.1 is assumed to be nonnegative and the map (4) is specified as $\hat{\pi} : k[x_1, \ldots, x_n] \to k[\theta_1, \ldots, \theta_d]$ (see (Pachter & Sturmbfels 2005)). As described later in this paper, in the algebraic descriptions of maximum entropy models one has to deal with the Laurent polynomials and hence one has to include the negative case in the definitions of toric ideals and toric models. This poses no problem because toric ideal theory in commutative algebra naturally includes the negative case (as in Definition 3.1) and Gröbner bases theory can be extended to Laurent polynomial ring (Pauer & Unterkircher 1999).

The concept of toric ideals let to the description of exponential models under the name toric models in algebraic statistics which is defined as follows.

**Definition 3.2.** *Let $A \in \mathbb{Z}_{\geq 0}^{d \times m}$ be a matrix such that the vector $(1, \ldots, 1) \in \mathbb{Z}_{\geq 0}^{m}$ is in the row span of $A$. Let $h \in \mathbb{R}_{>0}^{m}$ be a vector of positive real numbers. Let $\Theta = \mathbb{R}_{>0}^{m}$ and let $\kappa^{A,h}$ be the rational parametrization*

$$\kappa^{A,h} : \Theta \to \mathbb{R}^m$$
$$\kappa_j^{A,h} : \theta \mapsto Z(\theta)^{-1} h_j \prod_{i=1}^{d} \theta_i^{a_{ij}} \ , \tag{8}$$

*where $\theta = (\theta_1, \ldots, \theta_d)$ and $Z(\theta)$ is the appropriate normalizing constant. The toric model is the parametric algebraic statistical model*

$$\mathcal{M}_{A,h} \triangleq \kappa^{A,h}(\Theta) \ . \tag{9}$$

Independence models, exponential models, Markov chains and Hidden Markov chains can be given an algebraic statistical description by means of toric models (Pachter & Sturmbfels 2005). We keep positivity of $A$ in the Definition 3.2 as a matter of convention.

*3.2. ME in terms of Toric Models*

Let $X$ be a random variable taking values from the set $[m] = \{1, 2, \ldots m\}$. The only information we know about the pmf $p = (p_1, \ldots, p_m)$ of $X$ is in the form of expected values of the functions $t_i : [m] \to \mathbb{R}$, $i = 1, \ldots, d$ (we refer these functions as 'constraint functions'). We therefore have

$$\sum_{j=1}^{m} t_i(j) p_j = T_i \ , i = 1, \ldots d \ , \tag{10}$$

where $T_i$, $i = 1, \ldots, d$, are assumed to be known. In an information theoretic approach to statistics, known as Jayens maximum entropy model, one would choose the pmf $p \in \Delta_{m-1}$ that maximize the Shannon entropy functional

$$S(p) = -\sum_{j=1}^{m} p_j \ln p_j \tag{11}$$

with respect to the constraints (10).

The corresponding Lagrangian can be written as

$$\Xi(p, \xi) \equiv S(p) - \xi_0 \left( \sum_{j=1}^{m} p_j - 1 \right) - \sum_{i=1}^{d} \xi_d \left( \sum_{j=1}^{m} t_i(j) p_j - T_i \right) \tag{12}$$

Holding $\xi = (\xi_1, \ldots, \xi_d)$ fixed, the unconstrained maximum of Lagrangian $\Xi(p, \xi)$ over all $p \in \Delta_{m-1}$ is given by an exponential family (Cover & Thomas 1991)

$$p_j(\xi) = Z(\xi)^{-1} \exp\left( -\sum_{i=1}^{d} \xi_i t_i(j) \right), j = 1, \ldots, m, \tag{13}$$

where $Z(\xi)$ is normalizing constant given by

$$Z(\xi) = \sum_{j=1}^{m} \exp\left( -\sum_{i=1}^{d} \xi_i t_i(j) \right). \tag{14}$$

For various values of $\xi \in \mathbb{R}^d$, the family (13) is known as *maximum entropy model*.

Now, we have following proposition.

**Proposition 3.3.** *The maximum entropy model (13) is a toric model provided that the constraint functions are integer valued.*

*Proof.* Set $\xi_i = -\ln \theta_i$, $i = 1, \ldots, d$. Now, (13) gives us

$$p_j = Z(\theta)^{-1} \exp\left( -\sum_{i=1}^{d} t_i(j) \ln \theta_i \right) = Z(\theta)^{-1} \prod_{i=1}^{d} \theta_i^{t_i(j)}. \tag{15}$$

By defining matrix $A = [t_i(j)] \in \mathbb{Z}^{d \times n}$ and setting $h = (\frac{1}{m}, \ldots, \frac{1}{m})$ we have rational parametrization as in (8). $\square$

Note that we allowed only integer valued functions in the ME-model in the above proposition, which is necessary for algebraic descriptions of the same. Here we also mention that in the above proof by assuming $h \in \Delta_{m-1}$ (which acts as a prior), we can imply that minimum I-divergence model (Csiszár 1975)

$$p_j = \widehat{Z}(\xi)^{-1} h_j \exp\left( -\sum_{i=1}^{d} \xi_i T_i(j) \right), j = 1, \ldots, m, \tag{16}$$

(with appropriate normalizing constant $\widehat{Z}(\xi)$) is indeed a toric model.

Once the specification of statistical model is done, the task is to calculate the model parameters with the available information. In this case the available information is in the form of expected valued of functions $t_i$, $i = 1, \ldots d$ and the Lagrange parameters $\xi_i$, $i = 1, \ldots, d$ are determined using the constrains (10).

## 4. Calculation of ME distributions via solving Polynomial equations

*4.1. Direct method*

One can show that the Lagrange parameters in ME-model (13) can be estimated by solving following set of partial differential equations (Jaynes 1968)

$$\frac{\partial}{\partial \xi_i} \ln Z(\xi) = T_i \ , i = 1, \ldots, d, \tag{17}$$

which has no explicit analytical solution. In literature there are several methods of estimating ME-models. One of the important method is Darroch and Ratcliff's generalized iterative scaling algorithm (Darroch & Ratcliff 1972). Here we can show that ME-models can be calculated using computational algebraic methods.

Note that set of all distributions which satisfies (10) is known as linear family (we denote this by $\mathcal{L}$). Now, if we represent the exponential family (13) by $\mathcal{E}$, the set of statistical models that results from ME-principle can be written as $\mathcal{L} \cap \mathcal{E} \subset \Delta_{m-1}$. One can show that $\mathcal{L} \cap \mathcal{E} \subset \Delta_{m-1}$ is a variety.

By substituting maximum entropy distributions (15) in (10) we get

$$\sum_{j=1}^{m} t_i(j) \prod_{i=1}^{d} \theta_i^{t_i(j)} = T_i Z(\theta) \ , \tag{18}$$

which can be written as

$$\sum_{j=1}^{m} t_i(j) \prod_{i=1}^{d} \theta_i^{t_i(j)} = T_i \sum_{j=1}^{m} \prod_{i=1}^{d} \theta_i^{t_i(j)} \ . \tag{19}$$

The solutions of system of polynomial equations (19) gives the maximum entropy model spcified the available information (10). We state this as a proposition.

**Proposition 4.1.** *The maximum entropy model (13) can be specified by solving set of polynomial equations provided that the constraint functions $t_i$, $i = 1, \ldots, d$ are integer valued.*

*4.2. Dual Method*

Here we follow the method of *dual* optimization problem. By using Kuhn-Tucker theorem we calculate Lagrange parameters $\xi_i$, $i = 1, \ldots, d$ in (13) by optimizing dual of $\Xi(p, \xi)$. That is the task is to find $\xi$ which maximizes

$$\Psi(\xi) \equiv \Xi(p^{(\xi)}, \xi) \ . \tag{20}$$

Note that $\Psi(\xi)$ is nothing but entropy of ME-distribution (13). We have

$$\Psi(\xi) = \ln Z + \sum_{i=1}^{d} \xi_i T_i \ . \tag{21}$$

This can be written as

$$\Psi(\xi) = \ln \sum_{j=1}^{m} \exp\left(-\sum_{j=1}^{d} \xi_i t_i(j)\right) + \sum_{i=1}^{d} \xi_i T_i$$

$$= \ln \sum_{j=1}^{m} \exp\left(\xi_i(T_i - t_i(j))\right) . \tag{22}$$

Now maximizing $\Psi(\xi)$ is equivalent to maximizing

$$\Psi'(\xi) = \sum_{j=1}^{m} \exp\left(\xi_i(T_i - t_i(j))\right) . \tag{23}$$

By introducing $\xi_i = -\ln \theta_i$, $i = 1, \ldots, d$ we have

$$\Psi'(\theta) = \sum_{j=1}^{m} \prod_{i=1}^{d} \theta_i^{t_i(j) - T_i} . \tag{24}$$

The solution is given by solving the following set of equations

$$\frac{\partial \Psi'}{\partial \theta_j} = 0 \ , j = 1, \ldots d. \tag{25}$$

Unfortunately $\frac{\partial \Psi}{\partial \theta_j} \in k[\theta_1^\pm, \ldots, \theta_d^\pm]$ only if $T_i \in \mathbb{Z}$. Now, we consider the case where the expected values are available as sample means.

In most practical problems the information in the form of expected values is available via sample or empirical means. That is, given a sequence of observations $O_1, \ldots, O_N$ the sample means $\widetilde{T}_i$, $i = 1, \ldots, d$, with respect to the functions $t_i$, $i = 1, \ldots, d$ are given by

$$\widetilde{T}_i = \frac{1}{N} \sum_{l=1}^{N} t_i(O_l), i = 1, \ldots, d, \tag{26}$$

and the underlying hypothesis is $T_i \approx \widetilde{T}_i$. That is

$$\sum_{j=1}^{m} p_j t_i(j) \approx \frac{1}{N} \sum_{l=1}^{N} t_i(O_l) \ , i = 1, \ldots, d. \tag{27}$$

Now we show that, by choosing alternate Lagrangian in the place of (12) we can transform the parameter estimation of ME-model to a problem of solving set of polynomial (Laurent) equations.

**Proposition 4.2.** *Given the hypothesis (27) the problem of estimating the ME-model in the dual method amounts to solving set of Laurent polynomial equations (assuming that constraint functions are integer valued).*

*Proof.* To retain the integer valued exponents in our final solution we consider the constrains of the form

$$N \sum_{j=1}^{m} t_i(j) p_j = \sigma_i \ , \quad i = 1, \ldots d \ , \tag{28}$$

where $\sigma_i = \sum_{l=1}^{N} t_i(O_l)$ denotes the sample sum. In this case Lagrangian is

$$\widetilde{\Xi}(p,\xi) \equiv S(p) - \xi_0 \left(\sum_{j=1}^{m} p_j - 1\right) - \sum_{i=1}^{d} \widetilde{\xi}_d \left(N \sum_{j=1}^{m} p_j t_i(j) - \sigma_i\right) . \quad (29)$$

This results in the ME-distribution

$$p_j(\xi) = \widetilde{Z}(\xi)^{-1} \exp\left(-N \sum_{i=1}^{d} \widetilde{\xi}_i t_i(j)\right) , \quad j = 1, \ldots, m, \quad (30)$$

where $Z(\xi)$ is normalizing constant given by

$$\widetilde{Z}(\xi) = \sum_{j=1}^{m} \exp\left(-N \sum_{i=1}^{d} \widetilde{\xi}_i t_i(j)\right) . \quad (31)$$

To calculate the parameters we maximize the dual $\widetilde{\Psi}(\widetilde{\xi})$ of $\widetilde{\Xi}(p,\xi)$. That is we maximize the functional

$$\widetilde{\Psi}(\widetilde{\xi}) = \ln \widetilde{Z} + \sum_{i=1}^{d} \widetilde{\xi}_i \sigma_i . \quad (32)$$

It is equivalent to optimizing the functional

$$\widetilde{\Psi}'(\widetilde{\xi}) = \sum_{j=1}^{m} \exp\left(\sum_{i=1}^{d} \widetilde{\xi}_i \sigma_i - N \sum_{i=1}^{d} \widetilde{\xi}_i t_i(j)\right)$$

By setting $\ln \widetilde{\theta}_i = \widetilde{\xi}_i$ we have

$$\widetilde{\Psi}'(\widetilde{\theta}) = \sum_{j=1}^{m} \prod_{i=1}^{d} \widetilde{\theta}_i^{(\sigma_i - N t_i(j))} \quad (33)$$

The solution is given by solving the following set of equations

$$\frac{\partial \widetilde{\Psi}'}{\partial \widetilde{\theta}_i} = 0 , \quad i = 1, \ldots d. \quad (34)$$

We have

$$\frac{\partial \widetilde{\Psi}'}{\partial \widetilde{\theta}_i} \in k[\widetilde{\theta}_1^{\pm}, \ldots, \widetilde{\theta}_d^{\pm}] , \quad i = 1, \ldots, d. \quad (35)$$

□

In algebraic statistics, algebraic descriptions are used to analyze the maximum likelihood estimates of exponential models (Pachter & Sturmbfels 2005). In the view that maximum likelihood and maximum entropy are related,it will be interesting to compare these two methods from algebraic statistical point of view.

*4.3. Kullback-Csiszar Iteration*

Minimum I-diverence princile is a generalization of maximum entropy principle, and which considers the cases where prior estimate of the distribution $p$ is available. Given a prior estimate $r \in \Delta_m$ and information in the form of (10) one would choose the pmf $p \in \Delta_m$ that minimizes the Kullback-Leibler divergence

$$I(p\|r) = \sum_{j=1}^{m} p_j \ln \frac{p_j}{r_j} \tag{36}$$

with respect to the constraints (10). The corresponding minimum entropy distributions are in the form of

$$p_j(\xi) = Z(\xi)^{-1} r_j \exp\left(-\sum_{i=1}^{d} \xi_i t_i(j)\right), j = 1, \ldots, m, \tag{37}$$

where $Z(\xi)$ is normalizing constant given by

$$Z(\xi) = \sum_{j=1}^{m} r_j \exp\left(-\sum_{i=1}^{d} \xi_i t_i(j)\right). \tag{38}$$

It is easy to see that estimating minimum entropy distributions can be translated to solving polynomial equations, when the feature functions are integer valued. Polynomial system one would solve in this case is

$$\sum_{j=1}^{m} r_j(t_i(j) - T_i) \prod_{i=1}^{d} \theta_i^{t_i(j)} = 0. \tag{39}$$

Hence we have following proposition.

**Proposition 4.3.** *The estimation of minimum entropy model (??) amounts to solving a set of polynomial equations in indeterminates $\theta_i = \exp(-\xi_i)$, $i = 1, \ldots, d$ provided that the feature functions $t_i$, $i = 1, \ldots, d$ are positive and integer valued.*

Since an estimation of ME-distributions involves solving a system of nonlinear equations, which become inefficient, one would employ a interative method where one would estimate the distibution considering only one constraint at a time. We describe this procedure as follows.

At $N^{\text{th}}$ iteration, the algorithm computes the distribution $p^{(N)}$ which minimizes $I(p^{(N)}\|p^{(N-1)})$ with respect the $i^{\text{th}}$ constraint, $1 \leq i \leq d$ if $N = ad + i$, for any positive integer $a$. In this iterative procedure we have $p^{(0)} = r$ and $p^{(1)}$ is given by

$$p_j^{(1)} = r_j \big(Z^{(1)}\big)^{-1} \zeta_1^{t_1(j)},$$

where $\big(Z^{(1)}\big)^{-1} = \sum_{j=1}^{m} r_j \zeta_1^{t_1(j)}$. Considering the first constraint in (10) can be estimated by soving polynomial equation

$$\sum_{j=1}^{m} r_j(t_1(j) - T_1) \zeta_1^{t_1(j)} = 0, \tag{40}$$

with inderminate $\zeta_1$. Similary we have
$$p_j^{(2)} = r_j \big(Z^{(1)}\big)^{-1} \big(Z^{(2)}\big)^{-1} \zeta_1^{t_1(j)} \zeta_2^{t_2(j)} \ ,$$
where $\big(Z^{(2)}\big)^{-1} = \sum_{j=1}^{m} \zeta_2^{t_1(j)}$. Considering the first two constrains in (10) ME distribution can be estimated by solving
$$\sum_{j=1}^{m} r_j(t_2(j) - T_2)\zeta_1^{t_1(j)} \zeta_2^{t_2(j)} = 0 \ , \tag{41}$$
along with (40).

In general, when $N = ad + i$ for some positive integer $a$, $p_j^{(N)}$, for $N = 1, 2 \ldots$ is given by
$$p_j^{(N)} = r_j \big(Z^{(1)}\big)^{-1} \ldots \big(Z^{(N)}\big)^{-1} \zeta_1^{t_1(j)} \ldots \zeta_N^{t_N(j)}$$
and is determined by the following system of polynomial equations
$$\left.\begin{array}{ll}
\sum_{j=1}^{m} r_j(t_1(j) - T_1)\zeta_1^{t_1(j)} & = 0 \ , \\
\sum_{j=1}^{m} r_j(t_2(j) - T_2)\zeta_1^{t_1(j)} \zeta_2^{t_2(j)} & = 0 \ , \\
\vdots & \\
\sum_{j=1}^{m} r_j(t_i(j) - T_i)\zeta_1^{t_1(j)} \zeta_2^{t_2(j)} \ldots \zeta_N^{t_i(j)} & = 0 \ .
\end{array}\right\} \tag{42}$$

## 5. Conclusion and Directions for Future research

In this paper we attempted to describe maximum (and hence minimum) entropy model in algebraic statistical framework. We showed that maximum entropy models are toric models when the constraint functions are assumed to be integer valued functions and the set of statistical models results from ME-principle is indeed an variety. In a dual estimation we demonstrated that when the information is in the form of empirical means, the calculation of ME-models can be transformed to solving set of Laurent polynomial equations. Work on computational algebraic algorithms for estimating ME-models are in progress. We hope that this will also shed light on possible interesting algebraic structures in information theoretic statistics.